\documentclass[conference]{IEEEtran}
\IEEEoverridecommandlockouts
\usepackage{cite}
\usepackage{amsmath,amssymb,amsfonts}
\usepackage{amssymb}
\usepackage{algorithm}
\usepackage{algorithmicx}
\usepackage{algpseudocode}
\usepackage{amsmath}
\usepackage{graphicx}
\usepackage{textcomp}
\usepackage{xcolor}
\usepackage{subfigure}
\usepackage{multirow}    
\usepackage{booktabs}    
\usepackage{float}

\def\BibTeX{{\rm B\kern-.05em{\sc i\kern-.025em b}\kern-.08em
    T\kern-.1667em\lower.7ex\hbox{E}\kern-.125emX}}
\begin{document}

\title{A Weighted Random Forest Based Positioning Algorithm for 6G Indoor Communications\\}

	\author{Yang Wu\textsuperscript{1}, Yinghua Wang\textsuperscript{2}, Jie Huang\textsuperscript{1,2}, Cheng-Xiang Wang\textsuperscript{1,2*}, Chen Huang\textsuperscript{2,1}
		\\
		\textsuperscript{1}National Mobile Communications Research Laboratory, School of Information of Science and Engineering, 
		\\Southeast University, Nanjing 210096, China.\\
		\textsuperscript{2}Pervasive Communication Research Center, Purple Mountain Laboratories (PML), Nanjing 211111, China.\\
		\textsuperscript{*}Corresponding Author: Cheng-Xiang Wang\\
		Email: wu\_yang@seu.edu.cn, wangyinghua@pmlabs.com.cn,\\
		\{j\_huang, chxwang\}@seu.edu.cn, huangchen@pmlabs.com.cn
	}
\maketitle
\begin{abstract}
Due to the indoor none-line-of-sight (NLoS) propagation and multi-access interference (MAI), it is a great challenge to achieve centimeter-level positioning accuracy in indoor scenarios. However, the sixth generation (6G) wireless communications provide a good opportunity for the centimeter-level positioning. In 6G, the millimeter wave (mmWave) and terahertz (THz) communications have ultra-broad bandwidth so that the channel state information (CSI) will have a high resolution. In this paper, a weighted random forest (WRF) based indoor positioning algorithm using CSI-based channel fingerprint feature is proposed to achieve high-precision positioning for 6G indoor communications. In addition, ray-tracing (RT) is used to improve the efficiency of establishing channel fingerprint database. The simulation results demonstrate the accuracy and robustness of the proposed algorithm. It is shown that the positioning accuracy of the algorithm is stable within 6~cm in different indoor scenarios with the channel fingerprint database established at 0.2~m intervals.
\end{abstract}


\section{Introduction}
Although the continuous development of global positioning system has allowed outdoor positioning with satisfactory accuracy, there are great difficulties in performing high-precision indoor positioning. There are strong attenuation of radio waves caused by the building material and structure, complex indoor scenarios, and high personnel flow. These factors result in serious NLoS propagation and MAI in indoor environments. Meanwhile, 6G wireless communications put forward indoor positioning accuracy of centimeters and response time of milliseconds~\cite{b1},\cite{b2}. To figure out these challenges, there are two ideas as below. 

The first one is summarized as geometric positioning method in this paper, i.e., using geometric principles based on channel characteristics to estimate the position. It contains power measurement method based on received signal strength (RSS), time measurement method based on time of arrival (ToA) or time difference of arrival (TDoA), and angle measurement method based on angle of arrival (AoA). This kind of method is generally vulnerable to the interference of NLoS and MAI, leading to the low positioning accuracy. 

The second is fingerprint-based positioning method (FPM), first proposed in~\cite{b3}. FPM generally contains two stages: offline and online. In the offline stage, FPM establishes the channel fingerprint database by means of the actual measurement or simulation of the signal data in the positioning area. In the online stage, FPM compares the measured signal data with the channel fingerprint database to estimate the actual position of the terminal.

Owing to the convenience and low cost of obtaining RSS, traditional FPM generally collects RSS as channel fingerprint feature~\cite{b4},\cite{b5}, but there is serious NLoS propagation indoors, resulting in the instability of RSS. Although preprocessing methods such as Kalman filtering have been used to alleviate this instability~\cite{b6}, it still fails to avoid large errors when the received signal fluctuates heavily. Therefore, the FPM based on RSS cannot satisfy the requirements of high-precision positioning either. Since the orthogonal frequency division multiplexing (OFDM) technology was used as the physical layer standard of IEEE 802.11a, CSI can be used to characterize the channel information of each subcarrier~\cite{b7}. As a signal feature of the physical layer, CSI has higher granularity and stronger stability than RSS and can truly reflect multipath information. However, the previous researches on CSI-based FPM just use its amplitude and phase information directly, which cannot give full play to the potential of CSI.

The positioning algorithm in the online stage is mainly divided into the deterministic algorithm represented by the weighted k-nearest neighbor (WKNN)~\cite{b8}, the probabilistic algorithm represented by Bayesian estimation~\cite{b9}, and the machine learning (ML) algorithm which is more popular recently. The accuracy of the Bayesian estimation depends heavily on whether the assumed probability distribution conforms to the true data distribution. As a simplest supervised ML algorithm, WKNN predicts the results generally fairly and accurately, but its stability is poor for it cannot reflect the implicit relationship between the fingerprints and coordinates. To solve this problem, more superior ML algorithms have been used for indoor positioning, including support vector machines (SVM)~\cite{b10}, random forest (RF), deep neural networks (DNN)~\cite{b11}--\cite{b14}, etc. However, SVM does not perform well on multi-classification problems and DNN relies on a large number of sample sets and is difficult to train.

RF, as an integrated ML algorithm based on decision tree, has an excellent performance and been widely researched in the fields of artificial intelligence, classification prediction, and data modeling~\cite{b15},\cite{b16}. It can handle high-dimensional data without data deletion or normalization and its training speed is fast, so RF is greatly suitable for multi-classification problems, such as indoor positioning. Therefore, in this paper, we propose WRF positioning algorithm using CSI-based channel fingerprint feature so that the positioning accuracy and speed can satisfy the the requirements of 6G. 

However, as a FPM, the proposed algorithm will still consume huge manpower, material resources, and time to build channel fingerprint database. To solve this problem, RT is applied. As a deterministic modeling method, RT can obtain the channel parameters quickly and accurately~\cite{b17}.

Above all, in this paper we focus on a high-precision indoor positioning algorithm which can satisfy the requirements of 6G. The main contributions are as below.

\begin{itemize}
\item CSI-based channel fingerprint feature is proposed and RT simulation is applied to analyze and demonstrate the feasibility and superiority of it. 
\item WRF-based positioning algorithm is proposed and RT simulation is applied to analyze and demonstrate the accuracy and robustness of it.
\end{itemize}

The remainder of this paper is organized as follows. Section~II briefly introduces the characteristics of 6G indoor channel and the theory of CSI, then puts forward a CSI-based channel fingerprint feature. Section~III proposes WRF-based positioning algorithm based on RF and WKNN. Section~IV provides the simulation results based on RT and demonstrates the predominant performance of the proposed algorithm. Finally, conclusions are drawn in Section~V. 

\section{CSI-Based Channel Fingerprint Feature}\label{Sec2}

\subsection{6G Indoor Channel Characteristics}
The 6G-oriented wireless channel will be a full-bands channel, mainly including sub-6 GHz, mmWave, THz, and optical bands. In this paper, we mainly focus on the mmWave and THz.

Since the wavelength of mmWave and THz is short and close to the roughness of an ordinary object surface, when electromagnetic wave is incident on such a plane, not only specular reflection but also a large amount of diffuse scattering will be generated. Whether specular reflection or diffuse scattering, each reflection brings a great loss. Thus in the mmWave and THz bands, it is difficult for multiple reflection paths to reach the minimum detectable power at the receiver, so the number of reachable paths is greatly reduced compared with the lower frequency bands. In general indoor communication scenarios, the transmitted paths are severely attenuated and hardly contribute to the received signal. Moreover, the diffracted paths are weak as well. As a result, mmWave and THz have a good propagation direction and the effective propagation paths are mainly line-of-sight (LoS)\cite{b18},\cite{b19}.

Combined with  mmWave and THz channel characteristics and based on RT technique, a unified multi-ray channel model of mmWave and THz bands is proposed in \cite{b20}. The channel impulse response of this model can be described as below
\begin{equation}
\begin{aligned}
h(t)&= \alpha_{\text {LoS }} \delta\left(t-\tau_{\text {LoS }}\right) \mathbb{I}_{\text {LoS }}+\sum_{p=1}^{N_{\text {Ref }}} \alpha_{\text {Ref }}^{(p)} \delta\left(t-\tau_{\text {Ref }}^{(p)}\right) \\
&+\sum_{q=1}^{N_{\text{Sca}}} \alpha_{\text {\text{Sca} }}^{(q)} \delta\left(t-\tau_{\text{Sca}}^{(q)}\right)+\sum_{u=1}^{N_{\text {Dif }}} \alpha_{\text {\text{Dif} }}^{(u)} \delta\left(t-\tau_{\text {\text{Dif} }}^{(u)}\right)
\end{aligned}\label{eq1}
\end{equation}
where ${{\mathbb{I} }_{\text{LoS}}}$ is the indicator function to determine whether the LoS path exists, 1 for the presence and 0 for the absence, ${{\alpha }_{\text{LoS}}}$, $\alpha _{\text {Ref}}^{(p)}$, $\alpha _{\text{Sca}}^{(q)}$, and $\alpha _{\text{Dif}}^{(u)}$ are the attenuation amplitudes of the LoS path, the $p$th reflection path, the $q$th scattering path, and the $u$th diffraction path, and${{\tau }_{LoS}}$, $\tau _{\text{Ref}}^{(p)}$, $\tau _{\text{Sca}}^{(q)}$, and $\tau _{\text{Dif}}^{(u)}$ represent the time delay of the LoS path, the $p$th reflection path, the $q$th scattering path, and the $u$th diffraction path, respectively. This model combines the LoS, reflection, scattering, and diffraction paths and the accuracy of this model is verified by experimental measurements (0.06--1 THz). 

After the Fourier transform is performed on \eqref{eq1}, the channel frequency response (CFR) is obtained, then we discretely sample it with $t=\left\lceil(\tau /{{T}_{s}})]+k\cdot {{T}_{s}}\right\rceil$, where ${{T}_{s}}$ denotes the sampling interval and $\left\lceil \centerdot  \right\rceil $ denotes the rounding function. Finally, we obtain the CSI, i.e., CSI is the discrete sampling form of CFR with different carrier frequencies of OFDM as frequency points. In OFDM, the CSI model is expressed as
\begin{equation}
Y=HX+\zeta\label{eq2} 
\end{equation}
where $X$ denotes the transmitted signal vector, $Y$ denotes the received signal vector, $\zeta$ denotes the Gaussian white noise vector, and $H$ is the CSI matrix.

\subsection{CSI Feature Estimation}

CSI can characterize multidimensional signal features, including AoA, propagation delay, and path loss. This can be obtained by super-resolution detection algorithms, such as multiple signal classification (MUSIC) estimation algorithm, spatial smoothing (SS) algorithm, etc. To overcome the defect of MUSIC that cannot distinguish coherent signals and SS that fails to address coherence in complex environments, a CSI-based forward-backward spatial smoothing algorithm is proposed for joint estimation of AoA and ToA. 

\subsection{Maximum Power Path Extraction}

As the analysis in the previous subsection, mmWave and THz have a severe indoor propagation loss. Therefore, it is a significant difference between the power of the different paths detected by the receiver. In this case, most of the rays reaching the receiver do not work on predicting the position. 

\begin{figure}[t]
\centering
\centering\subfigure[Total RSS]
{
    \begin{minipage}[b]{.46\linewidth}
        \centering
        \includegraphics[scale=0.275]{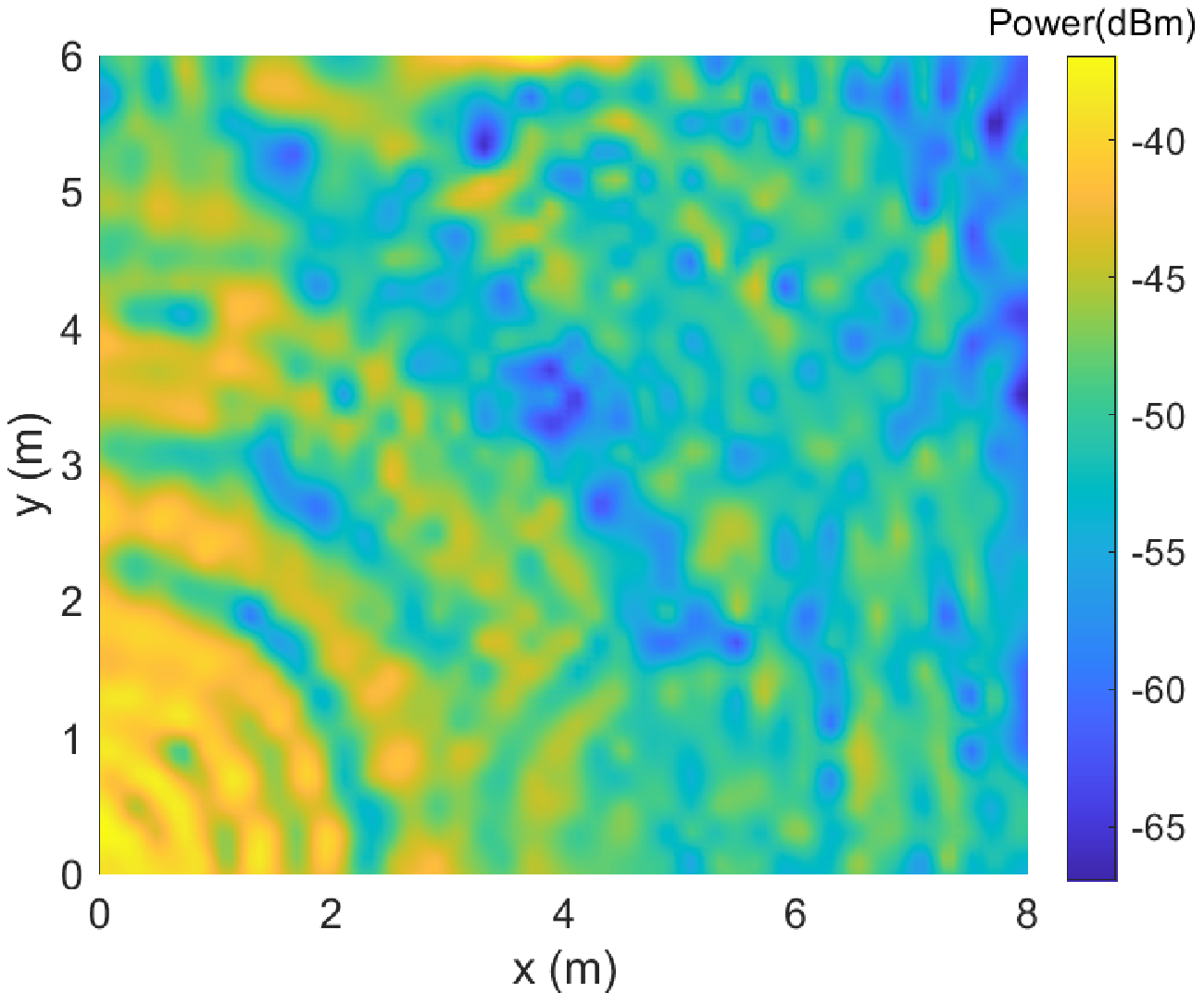}
        \label{fig1a}
    \end{minipage}
}
\centering\subfigure[RSS of MP path]
{
 	\begin{minipage}[b]{.46\linewidth}
        \centering
        \includegraphics[scale=0.275]{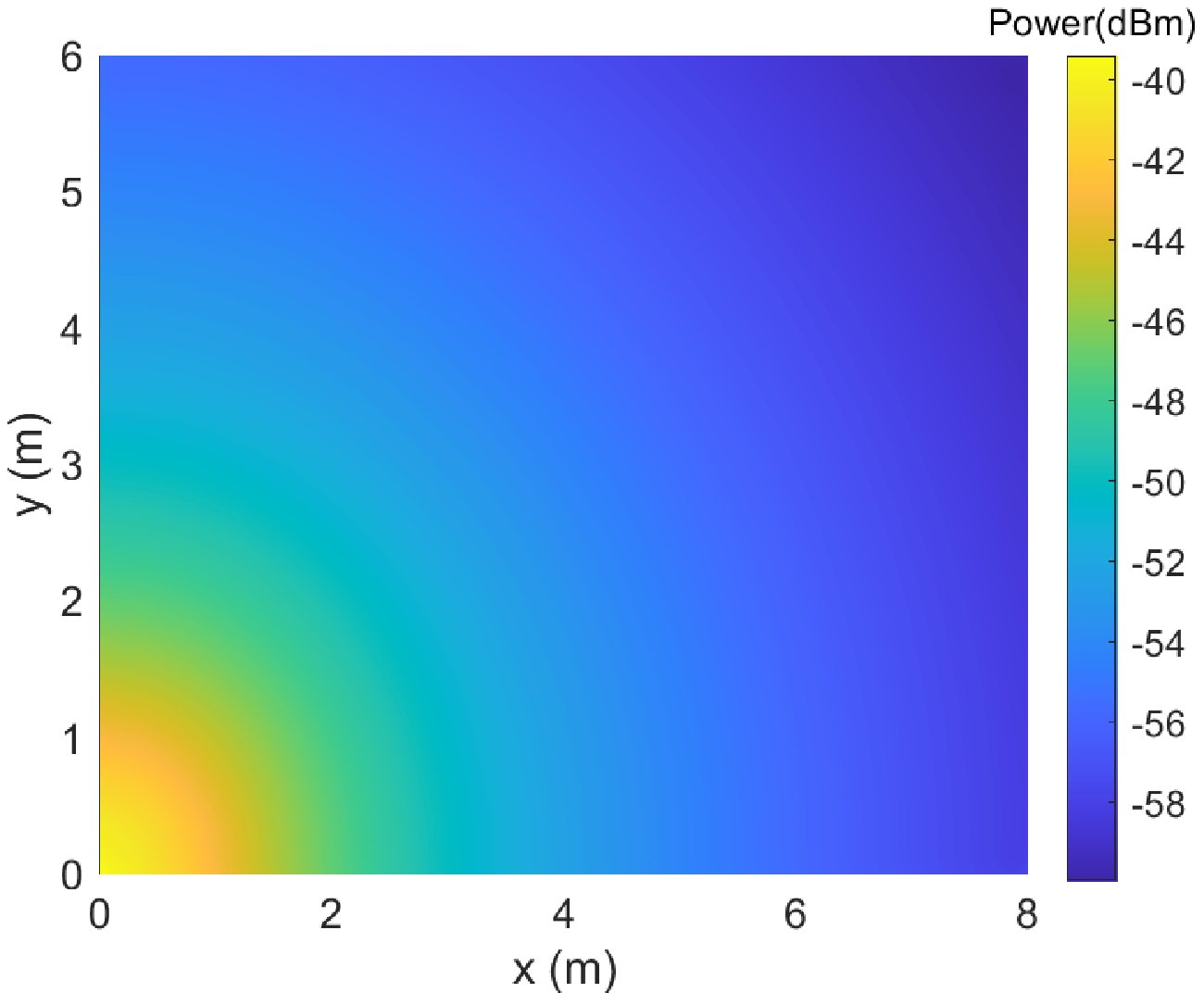}
        \label{fig1b}
    \end{minipage}
}
\caption{The spatial distribution of RSS with the transmitter at (0,0).}
\label{fig1}
\end{figure}

As we all know that the LoS path contributes the most to the received signal power with the highest energy among all the arriving rays. As a result, it plays a major role in position prediction. In addition, the LoS path power fits the path loss model best because its propagation path is not obscured by obstacles. However, due to the existence of severe NLoS propagation indoors, not every reference point (RP) has a LoS path. Thus in this paper, we propose the maximum power (MP) path. In indoor mmWave and THz communication scenarios, the MP paths at most RPs are the LoS paths. Even if the MP paths are NLoS paths, owing to the mmWave and THz characteristics, they only go through low-order reflection or diffraction, so their power loss is acceptable. Therefore, the RSS of MP path can fit the channel propagation loss model very well compared with the total RSS, as shown in Fig.~\ref{fig1}.

In addition, AoA and ToA have higher stability than RSS and can promote the distinction between fingerprint features. As a result, we integrate RSS, AoA, and ToA of MP path at the receiver as channel fingerprint features, i.e., CSI-based channel fingerprint features. Such channel fingerprint features not only improve fingerprint accuracy and distinction between RPs, but also remove redundant information and lessen the burden of the positioning algorithm.

\subsection{The Structure of Channel Fingerprint Database}

\begin{figure}[t]
\centering\includegraphics[scale=0.4]{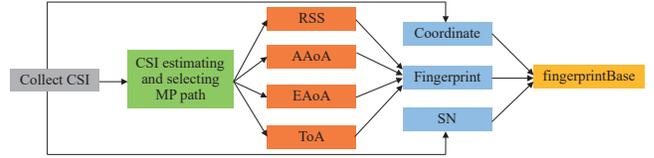}
\caption{The construction process and structure of fingerprintBase.}
\label{fig2}

\end{figure}

This subsection will introduce the structure of channel fingerprint database, as shown in Fig. \ref{fig2}. The channel fingerprint database is constructed as a structure named \textit{fingerprintBase}, which contains three elements: \textit{Fingerprint}, \textit{ Coordinate}, and \textit{SN}. SN is the AP number, a $1 \times n$ vector, and $n$ is the number of valid access points (APs). Coordinate is the RP coordinate, a $N \times 2$ matrix, and $N$ is the number of valid RPs. Fingerprint is the CSI-based channel fingerprint feature, a $N \times 4n$ matrix, containing RSS, azimuth AoA (AAoA), elevation AoA (EAoA), and ToA. In particular, it should be noted that, in order to adapt to the positioning algorithm later, the positioning area should be meshed perpendicular to the coordinate axis in the Cartesian coordinate system, then RPs are set at the grid intersections.

\section{Weighted Random Forest Positioning Algorithm}

\subsection{Weighted Prediction}

RF selects the result with the most votes in the decision tree as the prediction output. The higher the similarity, the more the votes, so RF is highly dependent on RPs. It is inevitable that there are many similar RPs in the channel fingerprint database, but just selecting the most similar one is obviously not comprehensive enough. The influence of other RPs on the positioning results is ignored, which leads to a large positioning error. In order to avoid that, a WRF algorithm is put forward. WRF combines the idea of WKNN algorithm, which selects the coordinates of the $K$ most similar RPs and uses the reciprocal of the similarity distance as the weight to predict the terminal coordinates. In the online stage, the fingerprint features extracted from test point (TP) are input into the model trained by RF and the output is no longer a single prediction coordinate, but the $K$ RPs' coordinates with the highest score $S$. The higher the $S$, the greater the impact on the estimation results, so the final predicted coordinates are calculated according to the score weighting of $K$ coordinates as follows

\begin{equation}
    {{(x,y)}_{\text{TP}}}=\frac{\sum\limits_{k=1}^{K}{({{x}_{k}},{{y}_{k}})\cdot {{S}_{k}}}} {\sum\limits_{k=1}^{K}{{{S}_{k}}}}.\label{eq3}
\end{equation}

\subsection{Coordinate Separation}

Although the joint coordinate estimation considers the uniqueness of each RP's fingerprint, it ignores the similarity of signal features among RPs. To further consider the similarity of signal feature between RPs, the training model $M\_{x}$ and $M\_{y}$ are derived by training the RPs' horizontal coordinates $x$ and vertical coordinates $y$ in the offline stage. In the online stage, the TP's channel fingerprint features are input into $M\_{x}$ and $M\_{y}$ to predict the $K$ horizontal coordinates $x$ and vertical coordinates $y$ with the highest score $S$. Then bring them into \eqref{eq3}

\begin{subequations}
\begin{align}
  {{x}_{\text{TP}}}={{\sum\limits_{k=1}^{K}{{{x}_{k}}\cdot S\_{{x}_{k}}}}} / {\sum\limits_{k=1}^{K}{S\_{{x}_{k}}}}\label{eq4a}\\
  {{y}_{\text{TP}}}={{\sum\limits_{k=1}^{K}{{{y}_{k}}\cdot S\_{{y}_{k}}}}} / {\sum\limits_{k=1}^{K}{S\_{{y}_{k}}}}\label{eq4b}.
\end{align}
\end{subequations}
The algorithm description of WRF is shown in Algorithm \ref{A1} and~\ref{A2}.

\subsection{The Framework of the Proposed Positioning Algorithm}
\begin{figure}[htbp]
\vspace{-0.5cm}
\centering\includegraphics[scale=0.47]{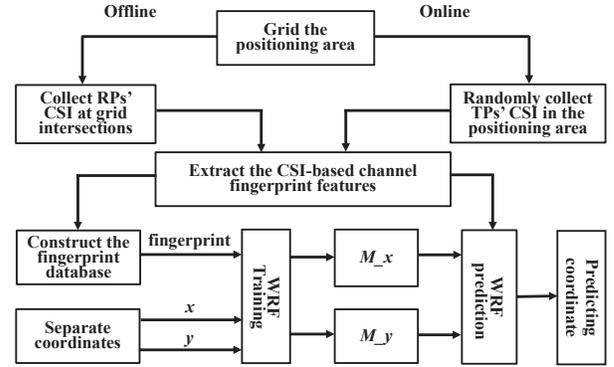}
\caption{The framework of the proposed positioning algorithm.}
\label{fig3}
\vspace{-0.2cm}
\end{figure}

In this subsection, we will combine the CSI-based fingerprint feature proposed in Section~\ref{Sec2} and the WRF algorithm proposed in this section to sort out the framework of the proposed positioning algorithm, as listed below and shown in Fig.~\ref{fig3}.

\begin{itemize}
\item Collect the CSI of offline RPs and online TPs. The RPs are located at the intersection of the grids established perpendicular to the coordinate axis in the Cartesian coordinate system.
\item Extract the CSI-based channel fingerprint features.
\item Construct the channel fingerprint database.
\item Separate the RPs' coordinates into $x$ and $y$ and send them into RF to be trained together with channel fingerprint features, then output the training model $M\_{x}$ and $M\_{y}$.
\item Send the online TPs' channel fingerprint features into $M\_{x}$ and $M\_{y}$ and carry out the weighted prediction, then output the final predicted TPs' coordinates.
\end{itemize}

\renewcommand{\algorithmicrequire}{\textbf{Input:}}
\renewcommand{\algorithmicensure}{\textbf{Output:}}
\begin{algorithm}[htbp]
\caption{WRF Training}
\begin{algorithmic}[1] 
\Require {1) The training sample set $T$ $\subset$ Fingerprint.} {2) The label set $L\_x$, $L\_y$ $\subset$ Coordinate.} {3) The number of Decision Tree $N$.}
\Ensure {The RF Classifiers $M\_x$, $M\_y$.}
\For{$L$ in \{$L\_x$, $L\_y$\}}
\State Initialize RF Classifiers $Model$ = $\varnothing$;
\State Randomly select $N$ sub-training sets $ST$ from $T$;
\For{$n$ in \{1, $\cdots$, $N$\}}
\State Initialize  Decision Tree $D(n)$ = $\varnothing$;
\State Randomly select features sets $\mathcal{F}$ in Fingerprint;
\Function {TreeGenerate}{$ST$, $\mathcal{F}$}
\State Generate node;
\If {All samples in $ST$ belong to $l$ in $L$}
\State {Mark this node as a leaf node of Label $l$;}
\State $D(n)$ = [$D(n)$ $\cup$ node];
\State \Return;
\EndIf
\If {$ST$  = $\varnothing$  \textbf{or} All samples in $ST$ have the same value on $\mathcal{F}$}
\State {Mark this node as a leaf node which is classed as the label with the most samples in $ST$;}
\State $D(n)$ = [$D(n)$ $\cup$ node];
\State \Return;
\EndIf
\State {Select the optimal feature ${f}$ whose value is ${f^{*}}$ from $\mathcal{F}$ according to Gini index;}
\State {Generate two branches for this node, the sample in $ST$ whose value is ${f^{*}}$ on the ${f}$ is marked as $D1$ and the rest are marked as $D2$;}
\State {\scshape{TreeGenerate}{($D1$, $\mathcal{F}$)}};
\State {\scshape{TreeGenerate}{($D2$, $\mathcal{F}$)}};
\EndFunction
\State $M$ = [$M$ $\cup$ $D(n)$];
\EndFor
\If{$L$ = $X$}
\State $M\_x$ = $M$;
\Else
\State $M\_y$ = $M$;
\EndIf
\EndFor
\State \Return $M\_x$, $M\_y$.
\end{algorithmic}
\label{A1}
\end{algorithm}

\renewcommand{\algorithmicrequire}{\textbf{Input:}}
\renewcommand{\algorithmicensure}{\textbf{Output:}}
\begin{algorithm}
\caption{WRF Estimating}
\begin{algorithmic}[1] 
\Require {1) The RF Classifiers $M\_x$, $M\_y$.} {2) The testing sample set $\mathcal T$ $\subset$ testFingerprint.} 
\Ensure {The predicting coordinate $(x_{TP}, y_{TP})$.}
\For{M in \{$M\_x$, $M\_y$\}}
\Function{RFPredict}{M, $\mathcal T$}
\For{$n$ in \{1,$\cdots$,$N$\}}
\State Compute the prediction $p_{n}$of the $D(n)$;
\EndFor
\State\textbf[$S$, $P$] = {Vote}{[$p_{1}$, $\cdots$, $p_{n}$]}
\EndFunction
\State Select the highest $K$ score $S$ and corresponding position $P$;
\If{$M$ = $M\_x$}
\State $x$ = $P$;
\State Compute the $x_{TP}$ according to \eqref{eq4a};
\Else
\State $y$ = $P$;
\State Compute the $y_{TP}$ according to \eqref{eq4b};
\EndIf
\EndFor
\State \Return $(x_{TP}, y_{TP})$.
\end{algorithmic}
\label{A2}
\end{algorithm}

\section{Analysis of RT Simulation Results}
\subsection{Simulation Parameters Setting}

In this section, we will construct the indoor scenarios and simulate the channel parameters with the commercial software Wireless InSite®. We set two different indoor scenarios. The area of scenario~1 is 16~m × 15~m and the area of scenario~2 is 8~m × 6~m, as shown in Fig.~\ref{fig5}. As well, the simulation parameters are shown in Table~\ref{tab1}. It should be pointed out that during the simulation, the reflection order is limited to 6 and the diffraction order is 1, which covers a reasonable portion of rays that could be detected in the real environments.

\begin{figure}[htbp]
\centering
\centering\subfigure[Scenario 1]
{
    \begin{minipage}[b]{.45\linewidth}
        \centering
        \includegraphics[scale=0.27]{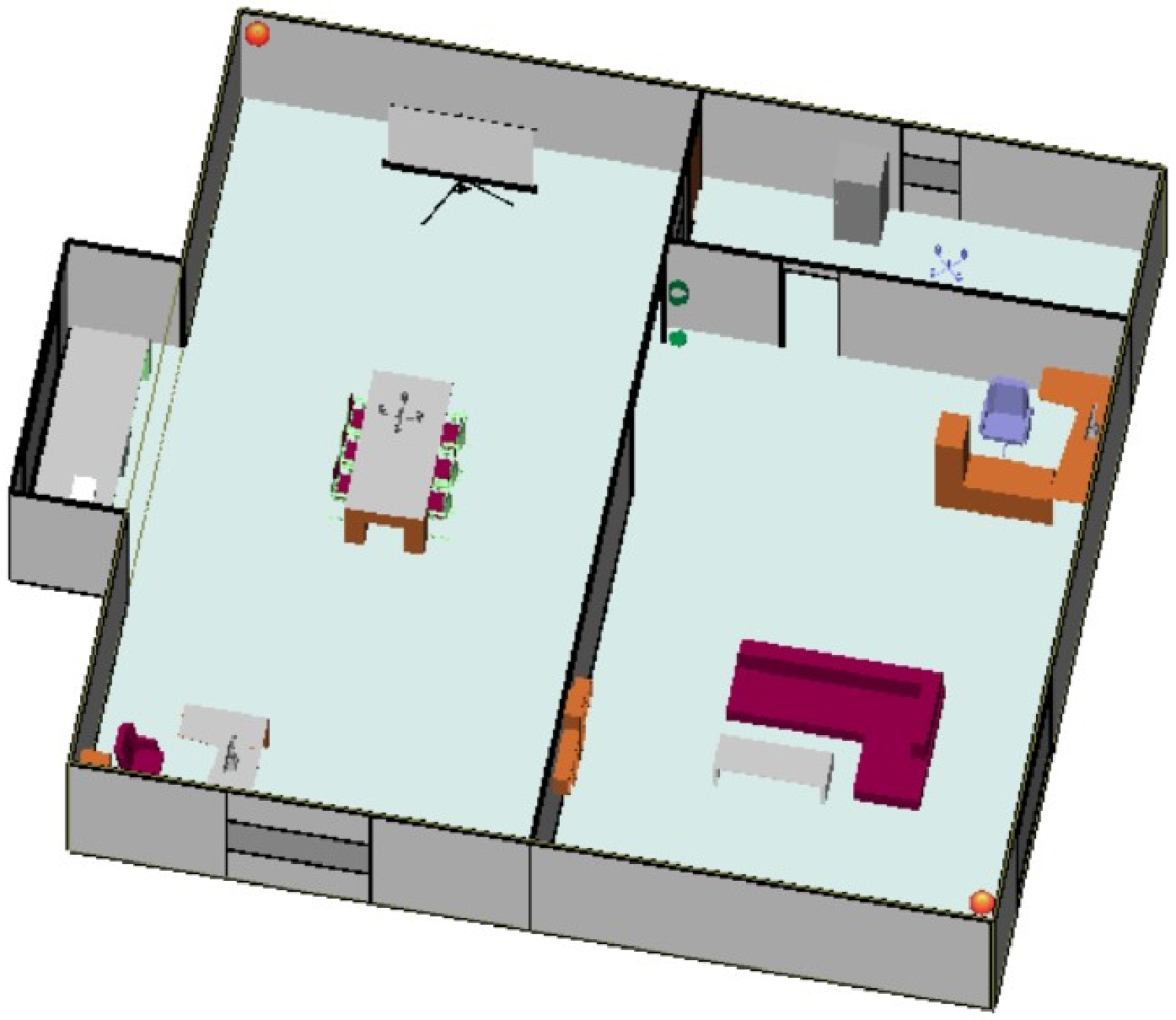}
        \label{fig4a}
    \end{minipage}
}
\centering\subfigure[Scenario 2]
{
 	\begin{minipage}[b]{.45\linewidth}
        \centering
        \includegraphics[scale=0.48]{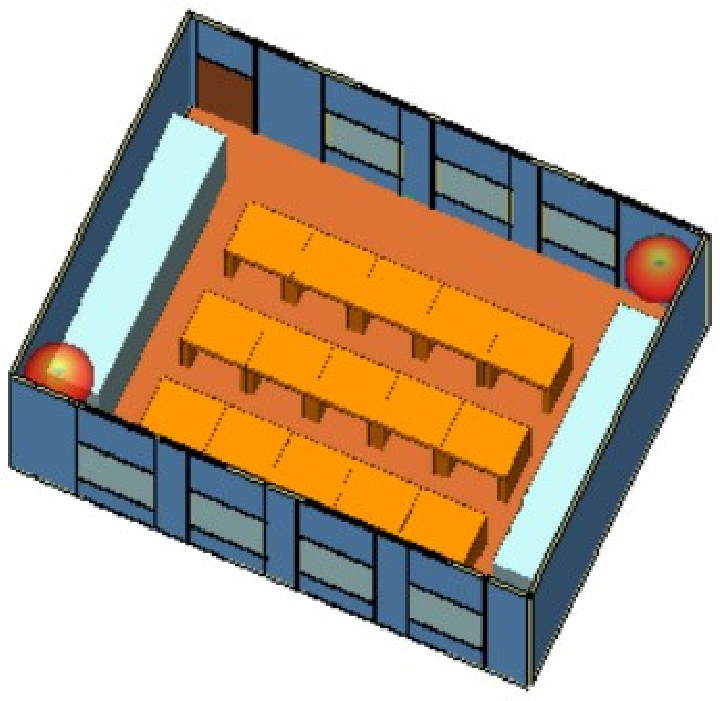}
        \label{fig4b}
    \end{minipage}
}
\caption{Simulation scenarios.}
\label{fig4}
\vspace{-0.5cm}
\end{figure}

\begin{table}[htbp]

\caption{The simulation parameters.}
\begin{center}
\setlength{\tabcolsep}{0.5mm}{
\renewcommand\arraystretch{1.1}    
\begin{tabular}{c c}
\toprule
{Simulation parameters}&{Value}\\
\midrule
{RPs' height}&{1.5 m}\\
\midrule
{APs' height}&{2.8 m}\\
\midrule
\multirow{2}*{APs' position}&{(-7.64 m, -6.72 m), (7.72 m, -7.54 m) in scenario 1 }\\
    &{(0.18 m, 0.20 m), (7.70 m, 5.78 m) in scenario 2}\\
\midrule
{Frequency}&{60 GHz}\\
\midrule
{Bandwidth}&{3 GHz}\\
\midrule
{RPs' antenna}&{Single omnidirectional}\\
\midrule
{APs' antenna}&{MIMO omnidirectional}\\
\midrule
{Reflection order}&{6}\\
\midrule
{Diffraction order}&{1}\\
\bottomrule
\end{tabular}}
\label{tab1}
\end{center}
\vspace{-0.5cm}
\end{table}

\subsection{Performance Evaluation of CSI-based Channel Fingerprint Feature}
\begin{figure}[bp]
\vspace{-0.5cm}
\centering\includegraphics[scale=0.53]{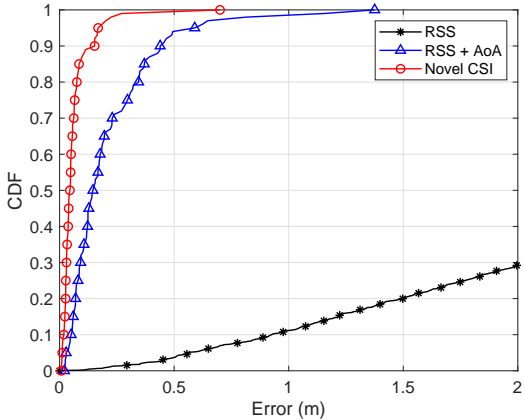}
\caption{CDF of positioning errors for different fingerprint features.}
\label{fig5}
\end{figure}

In this subsection, we construct the fingerprint database by setting reference grids with 0.2~m intervals and utilize WKNN algorithm to estimate the TPs' coordinates in the online stage. The collected fingerprint features are RSS, RSS $+$ AoA, and the proposed CSI-based channel fingerprint feature. The simulation results characterized by the error cumulative distribution function (CDF) are shown in Fig.~\ref{fig5}.

It can be seen that the probability of the error of the traditional RSS fingerprint feature within 1~m is only 10$\%$. After adding AoA as fingerprint feature, the accuracy is significantly improved. Further, the maximum error of the proposed CSI-based fingerprint feature is only 0.7~m and the average error is within 7~cm, which reaches the centimeter level. This is due to the fact that the proposed CSI fingerprint feature extracts the most fine-grained signal features of RPs and makes full use of the multipath information. Thus it can effectively resist the influence of NLoS propagation and MAI, resulting in significantly improved positioning accuracy.

\subsection{Performance Evaluation of WRF Algorithm}
\begin{figure}[t]
\vspace{-0.5cm}
\centering
\centering\subfigure[Scenario 1]
{
    \begin{minipage}[b]{.9\linewidth}
        \centering
        \includegraphics[scale=0.53]{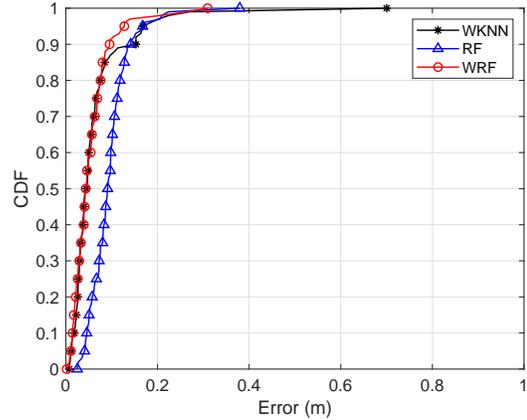}
        \label{fig6a}
    \end{minipage}
}
\centering\subfigure[Scenario 2]
{
 	\begin{minipage}[b]{.9\linewidth}
        \centering
        \includegraphics[scale=0.53]{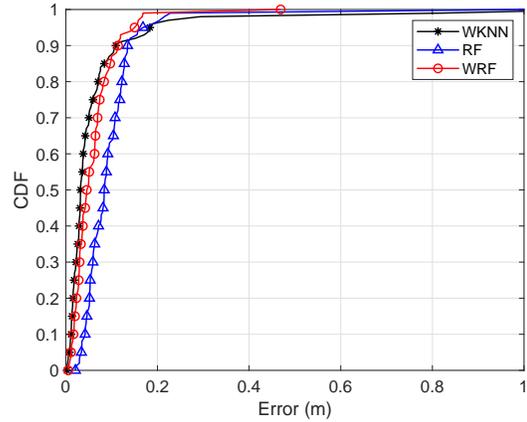}
        \label{fig6b}
    \end{minipage}
}
\caption{CDF of positioning errors for each algorithm with 0.2 m intervals.}
\label{fig6}
\vspace{-0.2cm}
\end{figure}

In this subsection, we mesh the Scenario~1 and Scenario~2 at intervals of 0.2~m, collect the proposed CSI-based fingerprint features as fingerprints to build fingerprint database, and set the number of WRF decision trees to 100. Then, we compare the WRF with RF\cite{b16} and WKNN\cite{b8} to verify the superiority and robustness of the WRF positioning algorithm. The simulation results are shown in Fig.~\ref{fig6} and Table~\ref{tab2}.

\begin{table}[t]
\caption{Performance of each algorithm with 0.2 m intervals in scenario~1.}
\begin{center}
\setlength{\tabcolsep}{0.4mm}{
\renewcommand\arraystretch{1.5}    
\begin{tabular}{|c|c|c|c|c|c|}
\hline
{Algorithm}&{Min(m)}&{Max(m)}&{Mean(m)}&{Training Time(s)}&{Positioning Time(s)}\\
\hline
{WKNN}&{0.0075}&{0.7002}&{0.0643}&{-}&{0.3}\\
\hline
{RF}&{0.0253}&{0.3796}&{0.0961}&{146.07}&{0.81}\\
\hline
\textbf{WRF}&\textbf{0.0018}&\textbf{0.3094}&\textbf{0.0552}&\textbf{23.64}&\textbf{0.74}\\
\hline
\end{tabular}}
\label{tab2}
\end{center}
\end{table}

It can be seen that the positioning error of WRF is smaller than WKNN and RF whatever in scenario~1 or in scenario~2. Although the error CDF of WRF are comparable with the first 80$\%$ of WKNN, the maximum error of WRF is only 0.3~m in scenario~1, which is much smaller than WKNN's 0.7~m, so WRF is much more stable than WKNN. Moreover, the average error of WRF is also the smallest, just about 5.5~cm, which reaches the centimeter level. Comparing from the positioning time, although the positioning time of WRF is longer than the WKNN, it still reaches millisecond level and is shorter than RF, especially its training time is significantly reduced (Note that Table~\ref{tab2} shows the positioning and training time of 100 TPs). In addition, WRF estimates the TPs' coordinates based on training model, so the positioning time changes less with the increase of database capacity. However, the online positioning time of WKNN will increase exponentially with the increase of database capacity, which can be seen from Table~\ref{tab3}. Therefore, with the increase of fingerprint database capacity, this difference between WRF and WKNN will decrease gradually.

\begin{table}[t]
\vspace{-0.5cm} 
\caption{Positioning time of each algorithm for different grid sizes.}
\begin{center}
{
\setlength{\tabcolsep}{3.5mm}{
\renewcommand\arraystretch{1.5}    
\begin{tabular}{|c|c|c|c|}
\hline
{Algorithm}&{WKNN}&{RF}&\textbf{WRF}\\
\hline
{1 m × 1 m}&{0.08 s}&{0.28 s}&\textbf{0.27 s}\\
\hline
{0.2 m × 0.2 m}&{0.3 s}&{0.81 s}&\textbf{0.74 s}\\
\hline
\end{tabular}}}
\label{tab3}
\end{center}
\vspace{-0.5cm} 
\end{table}

\section{Conclusions}
In this paper, a CSI-based WRF positioning algorithm has been proposed for 6G indoor communications and the performance of the algorithm has been demonstrated with RT simulation. The results have shown that the proposed algorithm has a maximum error within 0.3~m, an average error within 6~cm, and positioning time within 10~ms. This has indicated that the algorithm has very high accuracy, stability, and response speed. Thus, the proposed algorithm can satisfy the requirements of 6G indoor positioning with centimeter-level positioning accuracy and millisecond-level response time. In the future, we will carry out the realistic measurement for online TPs to further evaluate the proposed algorithm.

\section*{Acknowledgment}
This work was supported by the National Key R$\&$D Program of China under Grant 2018YFB1801101, the National Natural Science Foundation of China (NSFC) under Grants 61960206006 and 61901109, the Frontiers Science Center for Mobile Information Communication and Security, the High Level Innovation and Entrepreneurial Research Team Program in Jiangsu, the High Level Innovation and Entrepreneurial Talent Introduction Program in Jiangsu, the Research Fund of National Mobile Communications Research Laboratory, Southeast University, under Grant 2021B02, the EU H2020 RISE TESTBED2 project under Grant 872172, the High Level Innovation and Entrepreneurial Doctor Introduction Program in Jiangsu under Grant JSSCBS20210082, the Fundamental Research Funds for the Central Universities under Grant 2242022R10067, and the China Postdoctoral Science Foundation under Grant 2021M702499 and the Outstanding Postdoctoral Fellow Program in Jiangsu.

\end{document}